\documentclass[%
 reprint,
 amsmath,amssymb,
 aps,
]{revtex4-2}
\usepackage{amsmath,amsfonts,amssymb,mathtools}
\usepackage{hyperref}
\usepackage{graphicx}
\usepackage{color}
\usepackage{epsfig}
\usepackage{psfrag}	
\usepackage{tikz}
\usepackage{slashed}
\usepackage{tensor}
\usepackage{ulem}
\usepackage[font=small,labelfont=bf,width=1\textwidth]{caption}
\usepackage{comment}
\usepackage[all]{xy}
\usepackage{multirow}
\usepackage{float}
\usepackage{color}

\usepackage{ulem}
\usepackage{soul,xcolor}
\setstcolor{red}

\def\maybe#1{{\color [rgb]{0,.6,.6}}}

\DeclareMathOperator{\llangle}{\big\langle\hspace{-1.2mm}\big\langle\hspace{-.5mm}}
\DeclareMathOperator{\rrangle}{\hspace{-.5mm}\big\rangle\hspace{-1.2mm}\big\rangle}
\DeclareMathOperator{\Tr}{Tr}

\def\bF{{\mathbb{F}}}

\def\bC {\mathbb{C}}

\def\bO {\mathbb{O}}
\def\bP {\mathbb{P}}

\newcommand{\bea}{\begin{eqnarray}}
\newcommand{\eea}{\end{eqnarray}}
\newcommand{\beq}{\begin{equation}}
\newcommand{\eeq}{\end{equation}}
\newcommand{\bal}{\begin{equation}\begin{aligned}}
\newcommand{\eal}{\end{aligned} \end{equation}}

\newcommand{\vev}[1]{{\left< {#1} \right>}}
\newcommand{\vvev}[1]{{\left<\!\!\left< {#1}\right>\!\!\right>}}

\newcommand{\cG}{{\mathcal G}}

\newcommand{\cN}{{\mathcal N}}
\newcommand{\cP}{{\mathcal P}}

\newcommand{\cR}{{\mathcal R}}

\begin{document}

\title{Broken global symmetries and defect conformal manifolds}

\author{Nadav Drukker}
\email{nadav.drukker@gmail.com}
\author{Ziwen Kong}
\email{ziwen.kong@kcl.ac.uk}
\author{Georgios Sakkas}
\email{georgios.sakkas@kcl.ac.uk}
\affiliation{
Department of Mathematics, King's College London, The Strand, WC2R 2LS London, United Kingdom}

\begin{abstract}
Just as exactly marginal operators allow to deform a conformal field theory along 
the space of theories known as the conformal manifold, appropriate operators on conformal 
defects allow for deformations of the defects. When a defect breaks a global symmetry, there 
is a contact term in the conservation equation with an exactly marginal defect operator. 
The resulting defect conformal manifold is the symmetry breaking 
coset and its Zamolodchikov metric is expressed as the 2-point function of the 
exactly marginal operator. As the Riemann tensor 
on the conformal manifold can be expressed as an integrated 4-point function of the marginal 
operators, we find an exact relation to the curvature of the coset space.
We confirm this relation against previously obtained 4-point functions for insertions into 
the 1/2 BPS Wilson loop in $\cN=4$ SYM and 3d $\cN=6$ theory and the 1/2 BPS surface operator of 
the 6d $\cN=(2,0)$ theory.
    \end{abstract}

\maketitle

\section{Introduction and summary}
\label{sec:intro}

Amongst all operators of a conformal field theory (CFT), exactly marginal operators hold a special place, 
allowing for continuous deformations of the theory, forming a space of CFTs known as the 
conformal manifold. Those are common in supersymmetric theories, but otherwise not. In this note 
we point out that in the presence of conformal defects, one can define a similar notion of 
\emph{defect conformal manifold} and it naturally arises whenever a global symmetry is broken by the 
defect with supersymmetry or without.

Theories with conformal boundaries or defects are ubiquitous and play an 
important role both in condensed matter physics and in string theory. 
They form a defect CFT (dCFT) involving operators 
on and off the defect. A relatively unexplored topic (notable exceptions are 
\cite{Callan:1994ub, Recknagel:1998ih, Gaberdiel:2008fn,Behan:2017mwi, Karch:2018uft}) 
are marginal deformations of dCFTs by defect operators.

For a defect of dimension $d$, exactly marginal defect operators $\bO_i$ 
have scaling dimension $d$ and 
the correlation function of defect operators $\phi$
in the deformed theory can be expressed as
\begin{equation}
\label{deformedcor}
\llangle\phi\phi'\dots\rrangle_{\zeta^i}
=\llangle e^{-\int\zeta^i\bO_i\, d^dx}\phi\phi'\dots\rrangle_{0}\,.
\end{equation}
where $\zeta^i$ are local coordinates on the defect conformal manifold 
and the double bracket notation represents the correlation function in the dCFT normalized 
by the expectation value of the defect without insertions.

If the theory has a global symmetry $G$ with current 
$J^{\mu a}$, broken by the defect to $G'$, 
its conservation equation is modified to
\begin{equation}
\label{cons}
\partial_\mu J^{\mu a} = \bO_{i}(x_\parallel)\delta^{ia} \delta^{D-d}(x_\perp)\,,
\end{equation}
where $i$ is an index for the broken generators, $x_\parallel$ the directions along the defect and 
$x_\perp$ the transverse ones.

In a theory in $D$ dimensions, $J^{\mu a}$ has dimension $D-1$. 
Therefore $\bO_i$ has dimension $d$, so in the undeformed theory
\begin{equation}
\label{CO}
\vvev{\bO_i(x_\parallel)\,\bO_j(0)}=\frac{C_\bO \delta_{ij}}{x_\parallel^{2d}}\,.
\end{equation}
$C_\bO$ is fixed by the normalisation of $J^{\mu a}$ and determines the 
Zamolodchikov metric locally as 
$g_{ij}=C_\bO\delta_{ij}$~\cite{Zamolodchikov:1986gt}.

For $\phi=\phi'=\bO_i$, equation~\eqref{deformedcor} extends the Zamolodchikov metric 
beyond the flat space approximation. 
Differentiating~\eqref{deformedcor} with respect to $\zeta^i$ gives the Riemann tensor 
\cite{Kutasov:1988xb}
\begin{equation}
\begin{aligned}
\label{double-integral}
R_{ijkl}=\int d^dx_1 d^dx_2 
\Big[\llangle \mathbb{O}_j(x_1) \mathbb{O}_k(x_2) \mathbb{O}_i(0) \mathbb{O}_l(\infty)\rrangle_c
\\-\llangle \mathbb{O}_j(0)\mathbb{O}_k(x_2) \mathbb{O}_i(x_1) \mathbb{O}_l(\infty)\rrangle_c\Big]\,,
\end{aligned}
\end{equation}
where $\llangle\dots\rrangle_c$ indicates the connected correlator. 
This integral can be reduced to an integral over cross-ratios 
\cite{Friedan:2012hi}. See equations \eqref{friedan1d}, \eqref{ABJM-R2} \eqref{2dint} below.

When $\bO_i$ arise from symmetry breaking, the defect conformal manifold is the 
coset $G/G'$. It should be stressed that this statement 
is not evident in the correlation function calculation explained above, 
which is local on the conformal manifold. Rather it is a nontrivial statement 
that allows us to predict the form of the curvature (and higher derivatives 
of the metric).

Furthermore, the size of the coset is set by $C_\bO$ which appears in the metric and 
curvature tensor, making \eqref{double-integral} a non-trivial identity for integrated correlators. 
In the remainder of this paper we apply this idea in three examples: 1d dCFT of 1/2 BPS 
Wilson loops in $\cN=4$ supersymmetry Yang-Mills (SYM) in 4d and in the $\cN=6$ 
theory in 3d and the 2d dCFT of surface operators in the 6d $\cN=(2,0)$ theory. 
We derive explicit expressions for the Riemann tensor and verify it with known results for the 4-point functions.

\section{Maldacena-Wilson loops}
\label{sec:4dWL}

The 1/2 BPS Wilson loop along the Euclidean time direction in $\cN=4$ SYM is
\begin{equation}
\label{WL}
W=\Tr\cP e^{\int(iA_0+\Phi_6)dt}\,,
\end{equation}
The case of the 1/2 BPS circular loop has some subtle differences 
\cite{drukker:2000rr, Cuomo:2021rkm}, but here they are immaterial.

The defect CFT point of view on this observable was developed in 
\cite{Drukker:2006xg, correa:2012at, 
Drukker:2012de, Correa:2012hh, Gromov:2012eu, 
Liendo:2016ymz, cooke:2017qgm, Giombi:2017cqn, Liendo:2018ukf}. 
The lowest dimension insertions are the six scalar fields $\Phi_I$. 
Of them, $\Phi_6$ is marginally irrelevant, ``going up'' the 
renormalisation group flow to the UV non-BPS Wilson loop with no scalar coupling 
\cite{Alday:2007he, Polchinski:2011im, Giombi:2017cqn, Beccaria:2017rbe, 
Bruser:2018jnc, Grabner:2020nis, Cuomo:2021rkm}.

The remaining five scalars are to leading order $\bO_i$ of \eqref{cons}; 
the finite deformations being broken $SO(6)$ rotations
\begin{equation}
\Phi_6\to\cos\theta\,\Phi_6+\sin\theta\,\Phi_i\zeta^i/|\zeta|\,.
\end{equation}
It is natural to identify $|\zeta|=2\tan(\theta/2)$, extending the local 
metric in \eqref{CO} to the conformally flat metric on $S^5$
\begin{equation}
\label{global}
g_{ij}=\frac{C_\bO \delta_{ij}}{(1+|\zeta|^2/4)^2}\,.
\end{equation}

The 2-point function of $\Phi_i$ is indeed as in \eqref{CO} with 
$C_\Phi$ twice the bremsstrahlung function 
related to the expectation value of the circular Wilson loop 
\cite{Drukker:2011za, correa:2012at, Fiol:2012sg, Gromov:2012eu}
\begin{equation}
\begin{aligned}
\label{brem}
C_{\Phi}=\frac{1}{\pi^2}\lambda\partial_\lambda \log\vev{W_\circ}
=\frac{\sqrt\lambda}{2\pi^2}\frac{I_2(\sqrt\lambda)}{I_1(\sqrt\lambda)}+o(1/N^2)\,,
\end{aligned}
\end{equation}
where $\lambda$ is the 't~Hooft coupling and $I_n$ are modified Bessel functions.
At weak and strong coupling, 
this is
\begin{equation}
\label{CPhi-expand}
C_\Phi=
\begin{cases}
\frac{\lambda}{8\pi^2}- \frac{\lambda^2}{192\pi^2}+\frac{\lambda^3}{3072\pi^2}
-\frac{\lambda^4}{46080\pi^2}
+O(\lambda^5)\,,\\
\frac{\sqrt{\lambda}}{2\pi^2} -\frac{3}{4\pi^2} +\frac{3}{16\pi^2\sqrt{\lambda}}+\frac{3}{16\pi^2\lambda}+O(\lambda^{-3/2})\,,
\hskip1.5cm
\end{cases}
\hskip-2cm
\end{equation}

To express the 4-point function of $\Phi_i$, we define 
$\Phi^{(n)}=t_n^i\Phi_i(x_n)$ where $t^i_n$ are constant 5-vectors. 
Then~\cite{Liendo:2018ukf, Ferrero:2021bsb}
\begin{equation}
\begin{aligned}
\label{4pwithh}
{}&\llangle\Phi^{(1)}\Phi^{(2)}\Phi^{(3)}\Phi^{(4)}\rrangle
=C_\Phi^2\frac{t_{12}t_{34}}{x_{12}^2x_{34}^2}\cG(\chi;\sigma,\tau)\,,
\\
&\cG(\chi;\sigma,\tau)=\sigma h_2(\chi)
+\tau h_1(\chi)
+ h_0(\chi)
\,.
\end{aligned}
\end{equation}
with $t_{ij}\equiv t_i \cdot t_j$, $x_{ij}=x_i-x_j$ and the cross-ratios
\begin{align}
\label{cross-ratio}
\chi&=\frac{x_{12}x_{34}}{x_{13}x_{24}}\,, \quad
\\
\label{alpha}
\sigma&=\frac{t_{13} t_{24}}{t_{12}t_{34}}
=\alpha \bar{\alpha}\,,\quad 
\tau=\frac{t_{14} t_{23}}{t_{12} t_{34}}=(1-\alpha) (1-\bar{\alpha})\,.
\end{align}
The functions in \eqref{4pwithh} are fixed by superconformal symmetry to take the form
\begin{equation}
\begin{aligned}
\label{hFf}
h_0&=\chi^2\left(f/\chi-f/\chi^2\right)'\,,
\quad
h_1=-\chi^2(f/\chi)'\,,
\\
h_2&=\chi^2\mathbb{F}-\chi^2(f-{f}/{\chi})'\,,
\end{aligned}
\end{equation}
where $f$ is a function of $\chi$, prime is the derivative with respect to 
$\chi$ and $\mathbb{F}$ does not depend on $\chi$ and is determined 
from the topological sector of the correlators which occurs 
for the choice 
$\alpha=\bar\alpha=1/\chi$~\cite{Drukker:2009sf, Giombi:2018qox,Liendo:2018ukf}. 

Under crossing symmetry, $h_{0,1,2}$ transform as
\begin{equation}
\begin{aligned}
\label{crossingh}
&\chi^2h_2(1-\chi)=(1-\chi)^2h_2(\chi)\,,
\\
&\chi^2h_1(1-\chi)=(1-\chi)^2h_0(\chi)\,,
\\
&\chi^2h_0(1-\chi)=(1-\chi)^2h_1(\chi)\,.
\end{aligned}
\end{equation}

As the 4-point function depends only on the cross-ratio $\chi$, one can perform one of the 
two integrals in the curvature \eqref{double-integral} explicitly and reduce the formula 
to~\cite{Friedan:2012hi}
\begin{equation}
\begin{aligned}
\label{friedan1d}
R_{ijkl}=-\text{RV}\int_{-\infty}^{+\infty} &d\eta \log|\eta| 
\Big[ \llangle \Phi_i (1) \Phi_j (\eta) \Phi_k (\infty) \Phi_l(0)\rrangle_c 
\\&+\llangle \Phi_i (0)\Phi_j(1-\eta) \Phi_k(\infty) \Phi_l(1)\rrangle_c\Big]\,.
\end{aligned}
\end{equation}
RV denotes a particular prescription for regularizing and subtracting 
the divergences---a hard-sphere (point-splitting) cutoff 
followed by minimal subtraction~\cite{Friedan:2012hi}.

We can further reduce the integral to $\eta\in(0,1)$, but need to account for the subtlety that in 1d 
the order of the insertions is meaningful, so
\begin{equation}
\begin{aligned}
\label{order}
&\llangle \Phi_i(1)\Phi_j(\eta) \Phi_k (\infty)\Phi_l(0)\rrangle_c
\\&
=\begin{cases}
\llangle\Phi_j(\eta) \Phi_l(0) \Phi_i(1) \Phi_k (\infty) \rrangle\,,\quad \eta<0\, ,\\
\llangle \Phi_l(0) \Phi_j(\eta) \Phi_i(1) \Phi_k (\infty) \rrangle\,,\quad 0<\eta<1\, ,\\
\llangle \Phi_l(0) \Phi_i(1) \Phi_j(\eta) \Phi_k (\infty) \rrangle\,,\quad \eta>1\, .
\end{cases}
\end{aligned}
\end{equation}
To illustrate the calculation, we consider the contribution to \eqref{friedan1d} from 
the region $\eta \in(0,1)$. Using \eqref{4pwithh} and replacing $\eta$ with the cross-ratio 
$\chi$ \eqref{cross-ratio} we find
\begin{equation}
\begin{aligned}
-\int_{0}^{1}d\chi\bigg(
\frac{\log\chi}{\chi^2}\big[ g_{li}g_{jk}h_2(\chi)
+g_{lk}g_{ij}h_1(\chi)+g_{lj}g_{ik}h_0(\chi)\big]\\
+\frac{\log(1-\chi)}{\chi^2}\big[ g_{li}g_{jk}h_2(\chi)+g_{lj}g_{ik}h_1(\chi)
+g_{lk}g_{ij}h_0(\chi)\big]\bigg).
\end{aligned}
\end{equation}
Here we see all three tensor structures of bilinears of the metric, but after 
combining all three regions, the result must have the same tensor structure as a 
Riemann tensor. Finally using the crossing relations \eqref{crossingh} we find
\begin{equation}
\begin{aligned}
\label{R-as-g}
R_{ijkl}=2(g_{ik}g_{jl}-g_{il}g_{jk})
\int_0^1 \frac{d\chi}{\chi^2} \log\chi \big(h_2+h_1-2h_0\big)\, .
\end{aligned}
\end{equation}

Comparing with \eqref{4pwithh}, the integrand can be written as
\begin{equation}
\label{zetastar}
-\frac{2\log\chi}{\chi^2}\cG(\chi;\sigma^*,\tau^*)\,,
\quad
\sigma^*=\tau^*=-1/2\,.
\end{equation}
\eqref{R-as-g} has the structure of the curvature of the maximally symmetric space 
$S^5=SO(6)/SO(5)$ \eqref{global}. The integral in \eqref{R-as-g} is then related to 
the radius of the sphere.
In Appendix~A we simplify the integral to
\begin{equation}
\begin{aligned}
\label{Constraintf}
\int_{0}^{1} d\chi \left(\left(1-\frac{2}{\chi^3}\right)f
-\left(1+\frac{1}{\chi}\right)\mathbb{F}\right).
\end{aligned}
\end{equation}

$f$ and $\bF$ were calculated at strong coupling by explicit 
world-sheet Witten diagrams~\cite{Giombi:2017cqn} and extended up to fourth order 
in~\cite{Ferrero:2021bsb} based on the formalism in~\cite{Liendo:2016ymz, Liendo:2018ukf}. 
This results in
\begin{equation}
\label{SeriesofF}
\mathbb{F}=-\frac{3}{\sqrt{\lambda}}+\frac{45}{8}\frac{1}{\lambda^{3/2}}
+\frac{45}{4}\frac{1}{\lambda^2}+O(\lambda^{-5/2})\,.
\end{equation}
Writing $f$ in a power series
\begin{equation}
\label{Seriesoff}
f(\chi,\lambda)=\sum_{n=1}^{\infty}\lambda^{-\frac{n}{2}}f^{(n)}(\chi)\,,
\end{equation}
the first one is~\cite{Giombi:2017cqn}
\begin{equation}
\begin{aligned}
\label{f1}
f^{(1)}
&=-(1-\chi^2)\log(1-\chi)
\\&\quad
+\frac{\chi^3(2-\chi)}{(1-\chi)^2}\log(\chi)
-\frac{\chi(1-2\chi)}{1-\chi}\,.
\end{aligned}
\end{equation}
The integral in \eqref{Constraintf} can be computed for $f^{(1)}$ 
as well as for $f^{(2)}$, $f^{(3)}$, $f^{(4)}$ found in~\cite{Ferrero:2021bsb}, 
by integration by parts. We find for the Ricci scalar $R$ of \eqref{R-as-g}
\begin{align}
\label{SeriesofC}
\frac{R}{20}
=\frac{2\pi^2}{\sqrt{\lambda}}+\frac{3\pi^2}{\lambda}+\frac{15\pi^2}{4\lambda^{3/2}}
+\frac{15\pi^2}{4\lambda^2}+O(\lambda^{-5/2})\,.
\end{align}
This exactly agrees with the large $\lambda$ expansion of $1/C_\Phi$, whose inverse 
is in \eqref{CPhi-expand}, as expected for a sphere of radius $\sqrt{C_\phi}$.

The relation between the integrated 4-point function and $C_\Phi$ 
can also be deduced from the integral identities guessed in
\cite{Cavaglia:2022qpg}, as shown in Appendix~B. 
Checks against weak coupling expressions 
\cite{Kiryu:2018phb, Barrat:2021tpn, Cavaglia:2022qpg} were also 
performed there.

\section{1/2 BPS loop in 3d $\cN=6$ theory}

Another line defect with known 4-point function is the 
1/2 BPS Wilson loop of the $\cN=6$ theory in 3d~\cite{Drukker:2009hy, Bianchi:2020hsz}. 
The $SU(4)$ R-symmetry is broken by the defect to $SU(3)$, so the defect 
conformal manifold is $\bC\bP^3$. Now the marginal operators are chiral and 
have the structure of a supermatrix. The Zamolodchikov metric takes the form
\begin{equation}
g_{i\bar\jmath}=\vvev{\bO_i(0)\bar\bO_{\bar\jmath}(1)}=4B_{1/2}\delta_{ij}\,,
\end{equation}
where $B_{1/2}=\sqrt{2\lambda}/4\pi+\dots$ is the bremsstrahlung function for these operators 
\cite{Lewkowycz:2013laa, Bianchi:2017svd, Bianchi:2018scb}.

For the 4-point function we need to distinguish two orderings~\cite{Bianchi:2020hsz}
\begin{equation}
\begin{aligned}
\label{abjm4}
\vvev{\bO_i(x_1)\bar\bO_{\bar\jmath}(x_2)\bO_k(x_3)\bar\bO_{\bar l}(x_4)}
&=\frac{g_{i\bar\jmath}g_{k\bar l}K_1-g_{i\bar l}g_{k\bar \jmath}K_2}{x_{12}^2x_{34}^2}\,,
\\
\vvev{\bO_i(x_1)\bar\bO_{\bar\jmath}(x_2)\bar \bO_{\bar k}(x_3)\bO_{l}(x_4)}
&=\frac{g_{i\bar\jmath}g_{l\bar k}H_1-g_{i\bar k}g_{l\bar \jmath}H_2}{x_{12}^2x_{34}^2}\,.
\end{aligned}
\end{equation}
Here $x_1<x_2<x_3<x_4$ and $K_i$, $H_i$ depend on the 
cross-ratio $\chi$ \eqref{cross-ratio}. Other orderings can be determined by conformal 
invariance.

The curvature now splits according to chirality
\begin{equation}
\begin{aligned}
\label{ABJM-RIJKL}
R_{ij\bar k\bar l}&=(g_{i\bar l}g_{j\bar k}-g_{i\bar k}g_{j\bar l})\cR_1\,,
\\
R_{i\bar\jmath k\bar l}&=(g_{i\bar l}g_{k\bar\jmath}+g_{i\bar\jmath}g_{k\bar l})\cR_2\,,
\end{aligned}
\end{equation}
Plugging the expressions \eqref{abjm4} in \eqref{friedan1d} and accounting for the ordering 
\eqref{order}, we find
\begin{align}
\label{R1R2}
\cR_1&=\int_0^1\frac{d\chi}{\chi^2}\left[\log\frac{\chi}{1-\chi}(K_1+K_2)+2\log\chi(H_1+H_2)\right],
\nonumber\\
\cR_2&=\int_0^1\frac{d\chi}{\chi^2}\big[\log(1-\chi)(2H_1-2H_2-K_1)
\\&\hskip1.7cm{}
+\log\chi(2H_2+K_2)\big]\,.
\nonumber
\end{align}

The functions $H_i$ and $K_i$ are expressed in terms of functions $h(\chi)$ defined for 
$\chi\in(0,1)$ and $f(z)$ with $z=\chi/(\chi-1)<0$ as
\begin{equation}
\begin{aligned}
\label{Expressions}
H_1&=\chi^2(\chi(h/\chi)')'\,,
&
H_2&=\chi^2(\chi h')' \,,
\\
K_1&=z^2(z(f/z)')'
\,,
&
K_2&=z^2(z f')'
\,.
\end{aligned}
\end{equation}
In Appendix~C we show, based on crossing symmetry and assumptions on 
the behaviour of the functions in the limits $\chi\to0,1$, that $\cR_1=0$. 
Likewise, using~\eqref{R2-middle}, we find after repeated integration by parts
\begin{equation}
\label{ABJM-R2}
\cR_2=-2\int_0^1 \frac{h(\chi)}{\chi(1-\chi)}d\chi
+2\int _{-\infty}^0 \frac{f(z)}{(z-1)^2} dz\,.
\end{equation}

In~\cite{Bianchi:2020hsz} the functions $h$ and $f$ were evaluated at first order 
at strong coupling from the analytic bootstrap
\begin{equation}
\begin{aligned}
\label{h,f}
h^{(1)}&\propto-\frac{(1-\chi)^3}{\chi}\log (1-\chi)+ \chi(3-\chi) \log\chi +\chi -1\,, \\
f^{(1)}&\propto -\frac{(1-z)^3}{z}\log (1-z)+ z(3-z) \log|z| +z-1\,,
\end{aligned}
\end{equation}
with the same proportionality constant $1/(2\pi \sqrt{2\lambda})$, 
determined by explicit Witten diagram calculations. 
Evaluating the integral \eqref{ABJM-R2}, we find $\cR_2=\pi/\sqrt{2\lambda}$, 
so the Ricci scalar agrees to leading order with $12/2B_{1/2}$, as expected 
for $\bC\bP^3$. 
This calculation also serves as an independent derivation of the 
proportionality constant without relying on the $AdS$/CFT correspondence 
and could be used to determine further unknowns in higher loop calculations.

\section{Surface operators in 6d}
\label{sec:surface}

The 6d $\cN=(2,0)$ theory has 1/2 BPS surface operators~\cite{ganor:1996nf} with 
the geometry of the plane or the sphere. In the absence of a Lagrangian description, 
we cannot write an expression like \eqref{WL}, 
yet many properties of the surface operators are known. In particular, 
they carry a representation of 
the $A_{N-1}$ algebra of the theory~\cite{witten:1995zh,DHoker:2008rje, bachas:2013vza} 
and we focus on the fundamental representation, described by an M2-brane 
in $AdS_7\times S^4$~\cite{maldacena:1998im}. 

The defect CFT approach to surface operators was developed in~\cite{Drukker:2020atp}. 
In this case the scalar $\bO_i$ \eqref{cons} is associated to breaking of $SO(5)$ R-symmetry 
and is of dimension 2. 
As shown in~\cite{Drukker:2020atp}, the normalisation constant $C_\bO$ in \eqref{CO} is 
now related to the anomaly 
coefficients $c$ and $a_2$~\cite{graham:1999pm, Gentle:2015jma, Rodgers:2018mvq, 
Jensen:2018rxu, Estes:2018tnu,Chalabi:2020iie,Wang:2020xkc,Drukker:2020dcz} by
\begin{equation}
\label{c}
C_\bO=\frac{c}{\pi^2}=-\frac{a_2}{\pi^2}=\frac{1}{\pi^2}\left(N-\frac{1}{2}-\frac{1}{2N}\right).
\end{equation}

The curvature tensor \eqref{double-integral} is now written in terms of the 
complex cross-ratio~\cite{Friedan:2012hi}
\begin{equation}
\begin{aligned}
\label{2dint}
&R_{ijkl}=-2\pi \,\text{RV} \int d^2\eta \log|\eta|\llangle \mathbb{O}_i(1) \mathbb{O}_j(\eta) \mathbb{O}_k(\infty) \mathbb{O}_l(0)\rrangle_c\\
\end{aligned}
\end{equation}
We further simplify the integral by mapping $|\eta|>1$ to $|\eta|<1$ by a conformal 
transformation, giving
\begin{align}
\label{2dint2}
R_{ijkl}=-2\pi
\int_{|\eta|<1} 
\!\!\!\!
&d^2 \eta 
\log |\eta| \Big[\llangle \mathbb{O}_l(0) \mathbb{O}_{j} (\eta) \mathbb{O}_{i}(1) \mathbb{O}_k(\infty)\rrangle_c
\nonumber\\&
-\llangle \mathbb{O}_l(0) \mathbb{O}_{i} (\eta) \mathbb{O}_{j}(1) \mathbb{O}_k(\infty)\rrangle_c\Big] 
\,.
\end{align}
In this expression, $\eta$ is equal to the cross-ratio $\chi$ 
as defined in \eqref{crossratioUV}.

The analogue of \eqref{4pwithh}, \eqref{abjm4} is
\begin{equation}
\begin{aligned}
\label{4O}
&\llangle \mathbb{O}^{(1)} \mathbb{O}^{(2)}\mathbb{O}^{(3)} \mathbb{O}^{(4)}\rrangle
=C_\bO^2\frac{t_{12} t_{34}}{\vec{x}_{12}^4 \vec{x}_{34}^4} \mathcal{G}(\chi,\bar{\chi};\alpha,\bar{\alpha})\,,
\\
&\mathcal{G}=\sigma h_2(\chi,\bar\chi)
+\tau h_1(\chi,\bar\chi)
+ h_0(\chi,\bar\chi) +\mathcal{G}_2\,,
\end{aligned}
\end{equation}
where the cross-ratios are as in \eqref{alpha} and
\begin{equation}
\label{crossratioUV}
\begin{aligned}
U=\frac{\vec{x}_{12}^2 \vec{x}_{34}^2}{\vec{x}_{13}^2 \vec{x}_{24}^2}= \chi \bar{\chi}\,,
\quad V=\frac{\vec{x}_{14}^2 \vec{x}_{23}^2}{\vec{x}_{13}^2 \vec{x}_{24}^2}
= (1-\chi) (1-\bar{\chi})\,.
\end{aligned}
\end{equation}
The crossing equations for $h_i$ are as in \eqref{crossingh} but with $|\chi|^2$ and $|1-\chi|^2$.

$\cG_2$ in \eqref{4O} is parity odd, and using the symmetry of the integration domain in \eqref{2dint2}, 
it does not contribute to the curvature tensor. This is easy to verify for the expression in \eqref{G2}
by changing the integration variables to $U$, $V$. The same should hold to all orders.

The curvature tensor is then
\begin{equation}
\label{surf-final}
R_{ijkl}=-2\pi (g_{ik}g_{jl}-g_{il}g_{jk}) 
\int\limits_{|\chi|<1} d^2 \chi \frac{\log |\chi|}{|\chi|^2}
(h_0-h_2)\,.
\end{equation}
The integrand is in fact the parity even part of $\cG\log|\chi|/|\chi|^2$ with $\sigma^*=-1$ and $\tau^*=0$.

The 4-point function was calculated to first order at large $N$ from 
the M2-brane with the geometry of $AdS_3$ in $AdS_7$~\cite{Drukker:2020swu} 
resulting in
\begin{equation}
\begin{aligned}
h_0&=\frac{6}{N} U^2
\big[(V-U+1)\bar{D}_{3333}+U \bar{D}_{3322}-\bar{D}_{2222}\big]\,,\hskip-2mm
\\
h_1&=\frac{6}{N} U^2
\big[(U-V+1) \bar{D}_{3333}+\bar{D}_{3223}-\bar{D}_{2222} \big]\,.
\\
h_2&=\frac{6}{N} U^2
\big[ (U+V-1) \bar{D}_{3333}+\bar{D}_{3232}-\bar{D}_{2222} \big]\,,
\\
\label{G2}
\mathcal{G}_2&=-\frac{9}{2N}U^2 (\chi- \bar{\chi}) (\alpha -\bar{\alpha}) \bar{D}_{3333}\,.
\end{aligned}
\end{equation}
The $\bar D$ functions are given in~\cite{Arutyunov:2002fh}.
The expressions here are 16 times smaller than in~\cite{Drukker:2020swu} 
because of difference in the normalisation of 
$\bO_i$ compared to the $S^4$ coordinates $y_i$ in~\cite{Drukker:2020swu}.

By numerical integration, we confirm that
\begin{equation}
R_{ijkl}=\frac{N}{\pi^2}(\delta_{ik}\delta_{jl}-\delta_{il}\delta_{jk})\,,
\end{equation}
as expected for $S^4=SO(5)/SO(4)$ of radius squared 
$C_\bO\sim N/\pi^2$ in the large $N$ limit \eqref{c}.

\section{Discussion}
\label{sec:discussion}

The main result of this paper is that much like exactly marginal bulk operators, 
exactly marginal defect operators lead to defect conformal 
manifolds with Zamolodchikov metrics and with 
Riemann curvature given by an integrated 4-point function 
\eqref{double-integral}. 
In analogy to Goldstone's theorem~\cite{Nambu:1960tm,Goldstone:1961eq}, 
such marginal defect operators are guaranteed to exist when the 
defect breaks a global symmetry. 
Thus unlike bulk marginal operators, exactly marginal defect operators 
are ubiquitous.

We checked \eqref{double-integral} against 
known 4-point functions in three different examples and found a match with 
the curvature of the metric as in \eqref{global}. 
While these examples are in supersymmetric theories, 
symmetry breaking defects exist in many CFTs. For example, for 
the critical $O(N)$ model \cite{Wilson:1971dc} defects were studied 
in~\cite{PhysRevLett.84.2180,Allais:2014fqa, allais, ParisenToldin:2016szc, 
Cuomo:2021kfm} and our analysis applies there and possibly 
has experimental signatures.

Analogous constraints can be found for higher point functions (see e.g.~\cite{Barrat:2021tpn}). 
The fully integrated 
correlators are again derivatives of the Zamolodchikov metric, and 
therefore fixed by the metric of the manifold.

This integral constraint can be incorporated into bootstrap algorithms. 
This was implemented in the numerical analysis in~\cite{Cavaglia:2022qpg} 
leading to far improved accuracy. Likewise, 
it can be implemented in analytic studies, replacing the need for Witten 
diagram calculations in the 3d $\cN=6$ example~\cite{Bianchi:2020hsz} 
and extending it to higher orders~\cite{inprogress}.

The same analysis can be applied to Wilson loops and surface operators 
in higher dimensional representations, where a lot of the required calculations 
have already been done
\cite{Drukker:2005kx, Yamaguchi:2006tq, Hartnoll:2006is, Gomis:2006sb, Gomis:2006im, 
Chen:2007ir, Fiol:2013hna, Gentle:2015jma, Rodgers:2018mvq, Jensen:2018rxu, 
Estes:2018tnu,Chalabi:2020iie,Giombi:2020amn, Wang:2020xkc}.

Richer defect conformal manifolds can arise from less symmetric symmetry breaking 
than the examples discussed here. Such defects will have a variety of 
marginal operators with different 2-point functions and one could 
find integral constraints for different components of the Riemann tensor.

Defect conformal manifolds do not require broken 
symmetries. One natural setting is in 3d theories, where line operators are known to have 
multiple marginal couplings 
\cite{Drukker:2019bev, Correa:2019rdk, Drukker:2020opf, Agmon:2020pde, 
Drukker:2020dvr, Drukker:2022ywj}.
It is also natural to look at systems with both defect and 
bulk marginal operators to construct richer structures. Some work in that 
direction is in~\cite{Karch:2018uft}.

\acknowledgements
We are indebted to G. Bliard, S. Giombi, N. Gromov, C. Herzog, Z. Komargodski, C. Meneghelli, 
M. Probst, A. Stergiou, M. Tr\'epanier and G. Watts for 
invaluable discussions.
ND's research is supported by STFC grants
ST/T000759/1 and ST/P000258/1. 
ZK is supported by CSC grant No. 201906340174.
GS is funded by STFC grant ST/W507556/1.

\bigskip

\paragraph*{Appendix A. Simplifying the integral of $f(\chi)$:}
\label{simplificationR}

To simplify the integral in \eqref{R-as-g}, 
we plug in the expressions 
\eqref{hFf} to find
\begin{equation}
\begin{aligned}
\int_0^1 d\chi \log\chi\left(\chi\mathbb{F}-f
-\frac{2f}{\chi}+\frac{2f}{\chi^2}\right)'\,.
\end{aligned}
\end{equation}
Integrating by parts gives the boundary term
\begin{equation}
\begin{aligned}
\log\chi\left(\chi\mathbb{F}-f
-\frac{2f}{\chi}+\frac{2f}{\chi^2}\right)\bigg|_0^1\,.
\end{aligned}
\end{equation}
Noticing the boundary behaviour~\cite{Ferrero:2021bsb}
\begin{equation}
\label{limitf}
f(\chi)\sim\begin{cases} 
-\mathbb{F}\chi^2/2\,,&\chi\rightarrow 0\,,\\
\mathbb{F}/2\,,&\chi\rightarrow 1\,,
\end{cases}
\end{equation}
the only nonvanishing term is a divergence $\bF\log0$ which we 
express as an integral and combine with the result of integration by parts
\begin{equation}
\begin{aligned}
\label{RicciExp}
-
\int_0^1 d\chi \left(\frac{\bF}{\chi}+\mathbb{F}
-\frac{f}{\chi}-\frac{2f}{\chi^2}+\frac{2f}{\chi^3}\right).
\end{aligned}
\end{equation}
The crossing relation $\chi^2f(1-\chi)=-(1-\chi)^2f(\chi)$ 
leads to the integral identities
\begin{equation}
\begin{aligned}
\label{intid}
\int_0^1d\chi\,\frac{f}{\chi^2}=0\,,
\quad
\int_0^1d\chi\,\frac{f}{\chi}
=\int_0^1d\chi\,f\,.
\end{aligned}
\end{equation}
This finally allows us to further simplify \eqref{RicciExp} to \eqref{Constraintf}.

\paragraph*{Appendix B: Relation to integral identities of~\cite{Cavaglia:2022qpg}:}
\label{app:kolya}

A linear combination of the two integral constraints noticed in~\cite{Cavaglia:2022qpg} is
\begin{equation}\begin{aligned}
\label{constraint}
\int_0^1 d\chi\left( 3\frac{f}{\chi}-2\frac{\delta G}{\chi^2}(1+\log\chi)\right)
=\frac{1}{2C_\Phi}+3\mathbb{F}\,,
\end{aligned}
\end{equation}
where
$\delta G$, in the notations of Section~\ref{sec:4dWL}, is
\begin{equation}
\begin{aligned}
\frac{\delta G}{\chi^2}
=\mathbb{F}-\partial_{\chi} \left(\left(1-\frac{1}{\chi}+\frac{1}{\chi^2}\right)f\right).
\end{aligned}
\end{equation}

Using \eqref{limitf}, the left-hand side of \eqref{constraint} is
\begin{equation}
\begin{aligned}
2\mathbb{F}+\mathbb{F}\log\epsilon
+\int_0^1 &d\chi\bigg( 3\frac{f}{\chi}-2\mathbb{F}(1+\log\chi)
\\&-2\left(\frac{1}{\chi}-\frac{1}{\chi^2}+\frac{1}{\chi^3}\right)f\bigg).
\end{aligned}
\end{equation}
Then using \eqref{intid}, this is
\begin{equation}
\begin{aligned}
\int_0^1 d\chi\left(
\left(2-\frac{1}{\chi}\right)\mathbb{F}
+\left(1-\frac{2}{\chi^3}\right)f\right),
\end{aligned}
\end{equation}
and finally using our result for the integral \eqref{Constraintf}, this is the 
right-hand side of 
\eqref{constraint}, proving it.

\paragraph*{Appendix C: Integral identities for $H_i$, $K_i$:}
\label{app:abjm}
To simplify the integrals in \eqref{R1R2} we note that the crossing 
relation $(1-\chi)^2K_1(\chi)=-\chi^2K_2(1-\chi)$ implies
\begin{equation}
\begin{aligned}
\label{abjm-cint}
\int_0^1 \frac{d\chi}{\chi^2} \log\chi\, K_i
=-\int_0^1 \frac{d\chi}{\chi^2}\log(1-\chi)\, K_{3-i}\,,
\end{aligned}
\end{equation}
which yields
\begin{equation}
\cR_1=2\int_0^1\frac{d\chi}{\chi^2}\left[\log\frac{\chi}{1-\chi}K_1+\log\chi(H_1+H_2)\right].
\end{equation}
Using \eqref{Expressions}, this is a total derivative. Furthermore, 
assuming the asymptotics
\begin{equation}
\begin{aligned}
\label{hf-asymptotics}
h(\chi)&\sim
\begin{cases}
a_0\chi+a_1\chi\log\chi\,,&\chi\to0\,,\\
a_2(\chi-1)\,,&\chi\to1\,,
\end{cases}
\\
f(z)&\sim
\begin{cases}
b_0 z+b_1z\log|z|\,,&z\to0\,,\\
b_2+b_3\log|z|\,,&z\to -\infty\,,
\end{cases}
\end{aligned}
\end{equation}
we find
\begin{equation}
\label{int-H2}
\int_{0}^{1}\frac{d\chi}{\chi^2}\log\chi\,H_2=0
\end{equation}
and
\begin{equation}
\begin{aligned}
\cR_1&=2a_0-2b_0\,.
\end{aligned}
\end{equation}
This indeed vanishes in the perturbative analytic bootstrap, where $a_0=b_0=-b_2$
as a consequence of the crossing of $h$ and a braiding relation to $f$~\cite{Bianchi:2020hsz}.

By using \eqref{abjm-cint}, \eqref{hf-asymptotics} and \eqref{int-H2} we can also simplify the 
second line of \eqref{R1R2} to
\begin{equation}
\label{R2-middle}
\cR_2=2\int_0^1\frac{d\chi}{\chi^2}\log(1-\chi)(H_1-H_2-K_1)
\,.
\end{equation}

\raggedright

\bibliographystyle{apsrev4-1}
\bibliography{refs}
\end{document}